\documentclass{article}
\usepackage{graphicx,bm,doublespace}
\begin{document}
\title{The reflection-antisymmetric counterpart of the K\'arm\'an-Howarth dynamical equation}
\author{Susan Kurien \\
Center for Nonlinear Studies and Theoretical Division, \\ 
Los Alamos National Laboratory, Los Alamos, New Mexico} 
\date{\today}
\begin{spacing}{2}
\maketitle
\begin{abstract}
	We study the isotropic, helical component in homogeneous
	turbulence using statistical objects which have
	the correct symmetry and parity properties. Using these
	objects we derive an analogue of the K\'arm\'an-Howarth
	equation, that arises due to lack of mirror-reflection symmetry
	 in isotropic
	flows. The main equation we obtain is consistent with the
	results of O. Chkhetiani [JETP, 63, 768, (1996)] and
	V.S. L'vov et al. [http://xxx.lanl.gov/abs/chao-dyn/9705016,
	(1997)] but is derived using only velocity correlations, with
	no direct consideration of the vorticity or helicity. This
	alternative formulation offers an advantage to both
	experimental and numerical measurements. We also postulate,
	under the assumption of self-similarity, the existence of a
	hierarchy of scaling exponents for helical velocity
	correlation functions of arbitrary order, analogous to the
	Kolmogorov 1941 prediction for the scaling exponents of
	velocity structure function.
\end{abstract}

\section{Introduction}
In their 1938 paper on the statistical properties of homogeneous,
isotropic, reflection-symmetric turbulence, T. von K\'arm\'an and
L. Howarth derived the equation for the dynamics of the two-point
velocity correlation function \cite{Karm38}. This equation is of fundamental
importance since it relates the mean rate of change of energy to the flux of 
energy across a given correlation length $r$ in the flow. 
A form of this equation was used by A.N. Kolmogorov in 1941 \cite{Kolm41b} 
(K41) to derive one of
the few exact results known for isotropic, homogeneous, and 
reflection-symmetric turbulence, the ``4/5ths law'' which
relates the third-order longitudinal structure function to $\epsilon$,
the mean rate of energy dissipation
\begin{equation}
\langle ({\bf u}_L({\bf x}+{\bf r}) - {\bf u}_L({\bf x}))^3 \rangle = -{4\over5}
\epsilon r
\label{k41}
\end{equation}
where ${\bf u}_L$ is the component of the velocity along the separation
vector ${\bf r}$.
If the flow is not reflection-symmetric however, a new equation may
be derived to complement the K\'ar\'man-Howarth equation.
Three recent works have derived equations for third-order statistics in 
isotropic helical flows by considering velocity-vorticity correlations 
\cite{Chkh96,LPP97,GPP00}. In this paper, we show that the K\'arm\'an-Howarth
equation has a counterpart which arises due to parity-violation in isotropic
flows and which can be written solely in terms of two-point
velocity correlations. We demonstrate the equivalence of our result with 
those of \cite{Chkh96} and \cite{LPP97}.

We were motivated in this work by a series of investigations 
which proposed the use of the SO(3)
decomposition of tensor quantities, the structure functions, defined
by
\begin{equation}
S_{\alpha \beta}({\bf r}) = \langle (u_\alpha({\bf x+r}) - u_\alpha({\bf x}))
(u_\beta({\bf x+r}) - u_\beta({\bf x}))\rangle
\label{sfn2}
\end{equation}
in order to study the anisotropic contributions to their scaling.
The decomposition of the structure function into rotationally
invariant, irreducible subgroups of the SO(3) symmetry group
$S_{\alpha \beta}^{j=0}({\bf r}) + S_{\alpha \beta}^{j=1}({\bf r}) + ...$ allowed the
separation of the isotropic (indexed by $j=0$) from the anisotropic
(indexed by $j>0$) contributions to the structure function. This
procedure has allowed better quantification of the rate of decay of
anisotropy of the small scales in turbulence \cite {KS02,ADKLPS98,ABP99}.  
These analyses considered homogeneous, isotropic and
reflection symmetric flows. In the isotropic ($j=0$) sector, the
reflection symmetric structure function tensor has the form
\begin{equation}
S_{\alpha \beta}({\bf r}) = C_1(r)\delta_{\alpha \beta} +
C_2(r){r_\alpha r_\beta \over r^2}
\end{equation}
Homogeneity and incompressibility provide a constraint between the
scalar functions $C_1(r)$ and $C_2(r)$. 
If the condition of
reflection symmetry is dropped, there arises a further tensor
contribution to the isotropic sector given by $\epsilon_{\alpha \beta
\gamma} {r^\gamma \over r}$. This contribution is interesting because
it is isotropic (rotationally invariant), which implies that it
belongs in the $j=0$ sector, but is antisymmetric in ($\alpha,\beta$) and
changes sign under mirror reflection of ${\bf r}$. Since the second
order structure function is symmetric in its indices and does not
change sign under inversion of ${\bf r}$, it simply cannot to be used
to observe this antisymmetric contribution. In fact, when the
antisymmetric contribution is included in our decomposition, we are
effectively using the isotropic irreducible representation of the O(3)
symmetry group which includes operations that are not reflexion
invariant under ${\bf r} \rightarrow -{\bf r}$. Said differently, the 
elements $\Lambda$ of the orthogonal group O(3) satisfy 
$det(\Lambda)=\pm1$. The elements with determinant +1 form the SO(3) 
symmetry group of all (even-parity) rotations while those with 
determinant -1 are (odd-parity) reflections.
The present work demonstrates how to
access this isotropic, antisymmetric, odd-parity contribution using 
the tensor object
with the appropriate parity and symmetry properties. The dynamics of
such an object will provide the antisymmetric counterpart to the
K\'arm\'an-Howarth dynamical equation. 

In section 2, we present and discuss the second- and third-order
velocity correlations and their symmetric and antisymmetric
contributions.  In section 4 we derive the antisymmetric,
odd-parity counterpart of the K\'arm\'an-Howarth equation for
the second-order correlation function and show its equivalence to
previous results. In section 5 we postulate the existence of
generalized helical higher-order velocity correlations and their
scaling behavior under the assumption of self-similarity. Section 6
provides a summary and discussion.

\section{The symmetry and parity properties of the two-point velocity 
correlation functions}
\subsection{The second-order correlation tensor}
The two-point correlation tensor function of velocity 
fluctuations is defined by 
\begin{equation}
R_{\alpha \beta}({\bf r}) = \langle u_\alpha({\bf x}) u_\beta({\bf x+r}) \rangle
\label{rij}
\end{equation}
where ${\bf r}$ is the vector separation between two points, and
subscripts $\alpha, \beta$ are components in a chosen Cartesian
coordinate system. In homogeneous, isotropic, and not necessarily
reflection-symmetric turbulence, the correlation function may be
written as a sum of the dyadics \cite{Batc,Chan}
\begin{equation}
R_{\alpha \beta}({\bf r}) = A_1(r)\delta_{\alpha \beta} +
A_2(r){r_\alpha r_\beta \over r^2} + H(r)\epsilon_{\alpha \beta
\gamma}{r_\gamma \over r}
\label{rij_basis}
\end{equation}

Such a tensor may be written as the sum of its symmetric (in 
$\alpha, \beta$) and antisymmetric components as
\begin{eqnarray}
R_{\alpha \beta}({\bf r})&=& {R_{\alpha \beta}({\bf r})+R_{\beta
\alpha}({\bf r}) \over 2} + {R_{\alpha \beta}({\bf r})-R_{\beta
\alpha}({\bf r}) \over 2} \nonumber\\ &=& R_{\alpha \beta}^S({\bf r})
+ R_{\alpha \beta}^A({\bf r})
\end{eqnarray}
The symmetric contribution $R_{\alpha \beta}^S({\bf r})$ consists of the first two 
terms on the right side of Eq. (\ref{rij_basis}) while the antisymmetric 
contribution $R_{\alpha \beta}^A({\bf r})$ is the last term in Eq. (\ref{rij_basis}). 

If the flow is statistically homogeneous, then the 
incompressibility constraint implies
\begin{equation}
\partial_\alpha R_{\alpha \beta}({\bf r}) = \partial_\beta R_{\alpha \beta}({\bf r}) = 0
\end{equation}
where $\partial_\alpha (\cdot)$ denotes the partial derivative with
respect to $r_\alpha$. The incompressibility condition applies
separately to the symmetric and antisymmetric components as
$\partial_\alpha R_{\alpha \beta}^S({\bf r}) = \partial_\beta R_{\alpha \beta}^S({\bf r}) = 0$ and
$\partial_\alpha R_{\alpha \beta}^A({\bf r}) = \partial_\beta R_{\alpha \beta}^A({\bf r}) = 0$
since the symmetric and antisymmetric components are of opposite
parity. This is an interesting and useful property of these
correlation functions in the isotropic sector and for homogeneous
flows -- decomposition into symmetric and antisymmetric components
automatically separates the even- and odd-parity contributions. 

The symmetric part $R_{\alpha \beta}^S({\bf r})$ with tensor basis
as follows,
\begin{equation}
R_{\alpha \beta}^S({\bf r})= A_1(r)\delta_{\alpha \beta} +
A_2(r){r_\alpha r_\beta \over r^2}
\label{rijs_basis}
\end{equation}
has been analyzed extensively (see for example, \cite{Hinze}) under
the assumption of homogeneous, isotropic and mirror-symmetric
turbulence. These three conditions imply the translational, rotational
and reflectional invariance respectively of a given statistical
quantity used to describe the flow. Note that the structure function
(Eq. (\ref{sfn2})) is twice the symmetrized correlation function
$R^S_{\alpha\beta}$ plus twice the mean-square velocity
fluctuation. The latter addition makes the structure function galilean
invariant and hence a suitable candidate for the study of universal
statistics of the small scales.

The form of the antisymmetric tensor in the $j=0$ sector of the O(3)
representation is
\begin{eqnarray}
R_{\alpha \beta}^A({\bf r}) &=& \langle u_\alpha({\bf x})u_\beta({\bf x}+{\bf r})
\rangle - \langle u_\beta({\bf x})u_\alpha({\bf x}+{\bf r})\rangle \nonumber \\
&=&H(r)\epsilon_{\alpha \beta \gamma}{r^\gamma\over r}
\label{rija_basis}
\end{eqnarray}
Let us apply the incompressibility constraint to the antisymmetric
tensor form:
\begin{eqnarray}
\partial_\alpha(H(r)\epsilon_{\alpha \beta \gamma}r_{\gamma}/r) 
&=& \epsilon_{\alpha \beta \gamma}{r_\gamma \over r} \partial_\alpha
H(r) + \epsilon_{\alpha \beta \gamma}{H(r)\over r}(\delta_{\alpha
\gamma} - r_\alpha r_\gamma/r^2)\nonumber\\ &=& \epsilon_{\alpha \beta
\gamma} {r_\alpha r_\gamma \over r^2}{\partial H(r)\over \partial
r}\nonumber\\ &\equiv& 0
\label{hincomp}
\end{eqnarray}
In going from the second to the last lines of Eq. (\ref{hincomp}), we
have used the fact that contracting an antisymmetric tensor with a
symmetric one gives identically zero.  We conclude that
incompressibility does not provide any constraint on the scalar
coefficient $H(r)$ of the antisymmetric tensor contribution.

We can give an argument that the antisymmetrized correlation function
is galilean invariant by definition. Suppose we are in a frame moving
with velocity ${\bf U}$, then
\begin{equation}
R_{\alpha \beta}^A({\bf r}) = \langle (u_\alpha({\bf x})+{\bf
U})(u_\beta({\bf x}+{\bf r})+{\bf U})\rangle - \langle (u_\beta({\bf
x})+{\bf U})(u_\alpha({\bf x}+{\bf r})+{\bf U})\rangle
\end{equation}
It is seen that, because of homogeneity and the minus sign used
to antisymmetrize, any dependence on ${\bf U}$ drops out. Therefore,
we can hope that, as in the case of the structure functions, the
object $R_{\alpha \beta}^A({\bf r})$ will display the (universal)
properties of the small scales.

\subsection{The third-order correlation tensor}
Our goal is to derive the dynamical equation for the second-order
antisymmetric correlation $R_{\alpha \beta}^A({\bf r})$ as a counterpart 
to the K\'arm\'an-Howarth dynamical equation for the second-order
symmetric correlation $R_{\alpha \beta}^S({\bf r})$ (denoted in their paper
of 1938 \cite{Karm38} by $R_{i k}({\bm \xi})$). Since such an expression will 
involve the two-point third-order correlation function, we will first 
review its properties. 
\begin{equation}
S_{\alpha \gamma,\beta}({\bf r}) = \langle u_\alpha({\bf x}) u_\gamma({\bf x}) 
u_\beta({\bf x}+ {\bf r}) \rangle
\end{equation}
has the following properties in homogeneous turbulence. It is clearly
symmetric in indices ${\alpha,\gamma}$, with mixed symmetry in other
combinations ${\alpha, \beta}$ and ${\gamma,\beta}$ and in general of
mixed parity. By ``mixed'' we mean that the symmetry and parity
properties are indeterminate. 
In the isotropic tensor representation then, 
there are four terms
\cite{Hinze,Chkh96}
\begin{eqnarray}
S_{\alpha \gamma, \beta}({\bf r}) &=& S_1(r)\delta_{\alpha
\gamma}{r_\beta \over r} + S_2(r)(\delta_{\alpha \beta}{r_\gamma \over
r} + \delta_{\beta \gamma}{r_\alpha \over r}) + S_3(r){r_\alpha
r_\gamma r_\beta \over r^3} \nonumber\\ &+& S(r)(\epsilon_{\alpha
\beta \nu}{r_\nu r_\gamma \over r^2} + \epsilon_{\gamma \beta
\nu}{r_\nu r_\alpha \over r^2})
\label{sikj_tens}
\end{eqnarray}  
In anticipation of separating the terms of opposite symmetry as was
done in the case of the second-order correlation function, we write
\begin{eqnarray}
S_{\alpha \gamma, \beta}({\bf r}) &=& {S_{\alpha \gamma, \beta}({\bf
r})+ S_{\beta \gamma, \alpha}({\bf r}) \over 2}+ {S_{\alpha \gamma,
\beta}({\bf r})-S_{\beta \gamma, \alpha}({\bf r})\over 2}\nonumber\\
&=& S^S_{\alpha \gamma, \beta}({\bf r}) + S^A_{\alpha \gamma,
\beta}({\bf r})
\end{eqnarray}
where $S^A_{\alpha \gamma, \beta}$ is antisymmetric in $\alpha,\beta$ 
and has tensor contributions
as follows
\begin{eqnarray}
S^A_{\alpha \gamma, \beta}({\bf r}) &=& {\langle u_\alpha({\bf x})
u_\gamma({\bf x}) u_\beta({\bf x}+{\bf r})\rangle - \langle
u_\beta({\bf x}) u_\gamma({\bf x}) u_\alpha({\bf x}+{\bf r})\rangle
\over 2}\nonumber\\
&=&{S_1(r) - S_2(r)\over 2}\Big(\delta_{\alpha \gamma}{r_\beta \over r}-\delta_{\beta \gamma}{r_\alpha \over r}\Big) \nonumber\\
&+& {S(r)\over 2}\Big(2\epsilon_{\alpha \beta \nu}{r_\nu r_\gamma\over
r^2} + \epsilon_{\gamma \beta \nu}{r_\nu r_\alpha \over r^2} -
\epsilon_{\gamma\alpha\nu}{r_\nu r_\beta
\over r^2}\Big)
\label{sijka_basis}
\end{eqnarray}
These are the terms which were excluded in the
 K\'arm\'an-Howarth equation for reflection-symmetric flows.

\section{The antisymmetric component of the K\'arm\'an-Howarth equation} 
We now derive in a simple manner the dynamical equation for 
$R^A_{\alpha \beta}({\bf r})$. As in Hinze's \cite{Hinze} equation 1.48, 
starting from the Navier-Stokes equation
for homogeneous turbulence we can write the equation for $R_{\alpha \beta}$
\begin{equation}
{\partial \over \partial t} R_{\alpha \beta} - \partial_\gamma
S_{\alpha \gamma, \beta} + \partial_\gamma S_{\alpha,\gamma\beta}
=-{1\over \rho}(-\partial_\alpha K_{p,\beta} + \partial_\beta
K_{\alpha,p}) + 2\nu \partial_{\gamma \gamma} R_{\alpha \beta}
\label{hinze_rij}
\end{equation}
where $K_{\alpha,p} = \langle u_\alpha({\bf x})p({\bf x}+{\bf r})
\rangle$ and $p$ is the pressure.  We write a similar equation for
$R_{\beta \alpha}$ which we subtract from Eq. (\ref{hinze_rij}) and
divide throughout by 2.
\begin{eqnarray}
{\partial \over \partial t} \Big ({R_{\alpha \beta} -R_{\beta \alpha}\over
2}\Big) &-& \partial_\gamma \Big({S_{\alpha \gamma, \beta} -
S_{\beta\gamma,\alpha} \over 2}\Big) +
\partial_\gamma \Big({S_{\alpha,\gamma\beta} - S_{\beta,\gamma\alpha}\over 2}\Big) \nonumber\\
&=& {1\over 2}\Big(-{1\over \rho}(-\partial_\alpha K_{p,\beta} +
\partial_\beta K_{\alpha,p}) + {1\over \rho}(-\partial_\beta
K_{p,\alpha} + \partial_\alpha K_{\beta,p})\Big)
\nonumber\\
&+& 2\nu \partial_{\gamma \gamma}\Big({R_{\alpha \beta} - R_{\beta
\alpha} \over 2}\Big)
\label{rij_anti}
\end{eqnarray}
The pressure terms may be shown to vanish identically using
homogeneity and  incompressibility and assuming regularity as $r\rightarrow0$, 
as in the reflection-symmetric, isotropic case \cite{Hill,Frisch}. The
homogeneity condition $S_{\alpha \gamma, \beta}({\bf r}) =
S_{\beta,\gamma\alpha}(-{\bf r})$ adds a further constraint, giving
\begin{equation} 
{\partial \over \partial t} R^A_{\alpha \beta} - 2 \partial_\gamma
S^A_{\alpha \gamma, \beta} = 2\nu \partial_{\gamma\gamma} R^A_{\alpha \beta}. 
\label{rij_anti_final}
\end{equation} 
This equation is the antisymmetric counter-part to the K\'arm\'an-Howarth
equation for the second-order correlation functions. All the
quantities in this equation are relatively easily measured in experiments 
and numerical simulations since
no velocity derivatives are involved in the correlation functions,
only the velocities themselves. 
Substituting in Eq. (\ref{rij_anti_final}) the tensor forms for the 
antisymmetric correlation functions (Eqs. (\ref{rija_basis}) and 
(\ref{sijka_basis})) we arrive at the dynamical relation relating the scalars 
$H(r)$ and $S(r)$
\begin{eqnarray}
{\partial \over \partial t} H(r) - \Big( 2{\partial S(r) \over \partial r} +
6{S(r) \over r}\Big) = 2\nu \Big({\partial^2 H(r) \over 
\partial r^2}
+ {2\over r}{\partial H\over \partial r} - {2\over r^2}H(r)\Big)
\label{chkh}
\end{eqnarray}

This equation was derived by Chkhetiani \cite{Chkh96}
using the dynamics of velocity-vorticity correlations. In the 
present derivation, we have arrived at the
conclusion without the need to directly consider vorticity or
helicity. We only used the O(3) tensor representation for the
correlation function in homogeneous, isotropic flows in which symmetry 
and parity properties are trivially separated. 

\subsection{Derivation of KH-helical scaling law}
We apply the curl operator to the second-order antisymmetrized
correlation function Eq. (\ref{rija_basis}), and obtain the leading
order behavior of $H(r) = {\mathcal H} r /3$ (see Eq. \ref{H_lead} and
associated details in the Appendix) where the mean helicity ${\mathcal
H} = \langle {\bf u}\cdot {\bm \omega} \rangle/ 2 $.  We now
substitute this leading order dependence of $H(r)$ back into the KH
law,
\begin{equation}
{\partial \over \partial t}\Big({{\mathcal H} r \over 3} + \dots\Big)
- \Big(2{\partial \over \partial r} + {6\over r}\Big)S(r) = 2\nu
\Big({\partial^2\over \partial r^2} + {2\over r} {\partial \over
\partial r} - {2\over r^2}\Big)H(r)
\end{equation}
Here, if we make the same assumption as \cite{Chkh96}, that the main
contribution to the time-derivative comes from the linear term with
the next order terms not changing in the inertial range, and neglect
the right-hand side in the limit as $\nu \rightarrow0$,
\begin{equation}
S(r) = {{h}\over 30} r^2
\label{rsq}
\end{equation}
where $h$ is the mean helicity dissipation rate.
This agrees with the scaling law derived in \cite{Chkh96}. (There is a 
difference of a factor of 1/2 in the definition of mean helicity between 
\cite{Chkh96} and the present work.) The assumption
made in deriving this law is that we have fully developed, freely $decaying$ 
turbulence. These are the same assumptions made by Kolmogorov in deriving
the 4/5ths law and the energy spectrum. It is with this assumption that 
the following holds \cite{mark}
\begin{equation}
{\partial\over \partial t}{\mathcal H} = \nu \langle(\partial_k
v_i)(\partial_k \omega_i)\rangle = { h}.
\label{hel_bal}
\end{equation}
If a driving force is introduced, additional terms arise in the helicity 
balance equation \ref{hel_bal} (for example $\langle f\cdot \omega \rangle$)  
which may not directly allow us to derive Eq. (\ref{rsq}). It is however, 
not unreasonable
to expect that Eq. (\ref{rsq}) will hold for the steady-state, forced 
high-Reynolds number case. An argument similar to that which Frisch 
\cite{Frisch} used to prove the 4/5ths law for the forced case, would have to 
be used. This aspect will not be covered in the present work.

In the Appendix we show that an alternative form of Eq. \ref{rsq} may be
derived in the form of the following pair of equations
\begin{eqnarray}
{\bf u}_L\cdot ({\bf u}_T \times {\bf u}'_T) = {1\over
15}h r^2 \label{LPP}\\
{\bf u}_T \cdot (({\bf u}_L \times {\bf u}'_T) 
+ ({\bf u}_T \times {\bf u}'_L)) = -{1\over30}h r^2
\nonumber
\end{eqnarray}
where the velocity vector has been separated into its longitudinal
(along the separation vector ${\bf r}$) and transverse components as
${\bf u} = {\bf u}_L + {\bf u}_T$. The un-primed velocities denote
their value at ${\bf x}$ and primed velocities denote their value at
${\bf x}+{\bf r}$.  The first line of Eq. \ref{LPP} is equivalent to
the so-called ``2/15ths law'' derived by L'vov et al \cite{LPP97} (see 
Appendix for more details).

\subsection{The scaling behavior of higher-order correlation 
functions} 
The antisymmetrized correlation
functions may be thought of as newly defined structure functions
appropriate for helical flows. In the second and third order
($R^A_{\alpha \beta}$ and $S^A_{\alpha \gamma, \beta}$ respectively)
we have shown that the antisymmetrized correlation functions are galilean
invariant, so that sweeping effects are eliminated as in the case of the
symmetric structure functions, and we may hope for universal properties 
for small $r$. For the third-order correlation, we have 
seen that the scaling in the inertial range is $\sim r^2$ (Eq. (\ref{rsq})). 
Let us now make the K41 assumption of self-similarity such that $S^A_{\alpha
\gamma, \beta} \sim (R^A_{\alpha \beta})^{3/2}$ we obtain the inertial 
range behavior for $R^A_{\alpha \beta} \sim r^{4/3}$. This corresponds
to an inertial range scaling of the (helicity) cospectrum
$\tilde E_{12}(k_3)\sim k^{-7/3}$. The $k_3$ denotes the wavenumber component
in the direction mutually orthogonal to $\alpha=1$, and $\beta=2$.
This estimate for the scaling of the cospectrum coincides with the
Lumley 1967 estimate \cite{Lum67} for the
(anisotropic sector $j=2$) shear-stress (Reynolds) cospectrum
$\tilde E_{12}(k_1)$. However, the present dimensional estimate for the
cospectrum $\tilde E_{12}(k_3)$ is due to the reflection symmetry breaking,
not due to the rotational symmetry breaking.

If we construct
$n$th-order antisymmetrized correlation functions with scaling
exponents $\xi_n$ in the inertial range, the self-similarity argument
dictates that $\xi_n={2n\over 3}$. This would be the helical counter
part to the K41 scaling prediction for the structure functions which
says that the $n$th-order structure functions have scaling exponents
$\zeta_n = {n \over 3}$. It is not at all clear that self-similarity is a
reasonable assumption to make even in the case of low-order helical 
statistics \cite{Fris02}. This conjecture may only hold in the case 
of the maximal helical cascade in which there is no joint cascade of energy
\cite{Eyink02}.

\section{Conclusion}
The understanding of helicity dynamics in three-dimensional flows is 
still evolving. 
It has been known for some time 
that helicity is conserved in the fluid equations in the inviscid limit 
\cite{Mof69}. The simultaneous existence of both helicity and energy 
cascades to the high-wavenumbers was first considered by A. Brissaud
et al. \cite{BFLLM73}. In that work, the scenario for a pure helicity
or maximally helical cascade was also proposed, in which energy
cascade to the small scales is blocked, giving rise to an energy
spectrum $E(k)\sim k^{-7/3}$. R.H. Kraichnan showed \cite{Kra73},
based on physical considerations, that the scenario of joint energy
and helicity cascades to the high-wavenumbers, with recovery of the
Kolmogorov energy spectrum $E(k)\sim k^{-5/3}$ is more likely. This
joint-cascade picture has subsequently been strengthened by
observations in numerical simulations \cite{BorOrs97} from which it
seems likely that the helicity injected at the large scales cascades
downscale, more or less passively transported by the energy
cascade. More recently, Ditlevsen and Giuliani show, both
theoretically \cite{DitGiu01a} and using shell-model calculations
\cite{DitGiu01b}, that at high-Reynolds numbers a joint cascade of
energy and helicity must exist in some range of wavenumbers. They
argue that for wavenumbers larger than this range the
reflection-symmetry is restored by the dominant helicity dissipation
term. Q. Chen et al. \cite{CCE02} have shown by means of helical-wave
decomposition of the velocity field, that the detailed transfer of
energy (and helicity) between helical-wave modes of opposite parity is
consistent with the existence of a joint cascade, with $-5/3$ scaling
for both energy and helicity spectra. They also confirm their
theoretical predictions using numerical simulations. Their analysis
disagrees with \cite{DitGiu01a,DitGiu01b} over the precise range
wavenumber over which these cascades exist, but nonetheless, both
works agree that for high Reynolds numbers, a joint cascade of energy
and helicity will coexist for some range of wavenumbers, with parity
restoration at sufficiently large wavenumbers. The present analysis is
also consistent with the joint cascade at high Reynolds numbers.  The
original K\'arm\'an-Howarth result and the helical version derived
here are not mutually exclusive. The former picks out the
reflection-symmetric part of the flow, while the latter picks out the
reflection-antisymmetric part. The two contributions are measured by
different quantities which allow for the $separation$ of the parity
and symmetry properties of the flow. It is not clear, atleast to this
author, whether the present formulation predicts that energy and
helicity cascades will coexist for $all$ scales (consistent with
\cite{CCE02}) or for only a certain range of scales (consistent with
\cite{DitGiu01a}) at Reynolds numbers high enough. This might have to
be left to empirical tests using experimental and direct numerical
simulations data. Thus far, only the shell-model simulations of
Ditlevsen and Giuliani \cite{DitGiu01b} and Biferale et al
\cite{BPT98} exist to guide our intuition as to the scaling ranges of
the cascades.

From the analysis of \cite{CCE02} it appears that if one of the
helical modes is blocked, which can easily be done in simulations and 
shell-model calculations, but
may not be possible in real flows, a pure helicity cascade will
develop in the remaining mode which blocks the energy cascade down to
small scales, and yields an energy spectrum $E(k) \sim k^{-7/3}$. In
this sense, the dimensional (self-similarity) argument of section 3
for the scaling exponent of the cospectrum $\tilde E_{12}(k_3)\sim
k_3^{-7/3}$ is consistent with the scenario of a pure helicity
cascade. Without speculating on the feasibility of physically
achieving such a purely helical cascade, we remark that the
$cospectrum$ $\tilde E_{12}(k_3)$, of the two orthogonal components
along the third orthogonal direction, is a fundamentally different
object than the energy spectrum and may well display entirely
different functional behavior. The helicity cospectrum, which should
be identically zero for homogeneous, isotropic, reflection symmetric
turbulence is a sensitive measure of reflection symmetry
breaking \cite{KST01} and the presence of helicity. The present work has 
shown that it arises from precisely that 
contribution to the second-order correlation which was excluded in the 
original isotropic, homogeneous, reflection-symmetric 
K\'arm\'an-Howarth equation.

A further new possibility suggested by this work is the construction
of antisymmetric higher-order (greater than 3) correlation
functions. Assuming each pair of indices of an $n$th order, two-point
velocity correlation function can be appropriately antisymmetrized as
has been done here for the second and third order cases, we may have a
new series of objects, which we will call helical
structure functions. These, along with the usual structure functions familiar
from studies of isotropic, reflection-symmetric flows, would form a complete 
set of statistical objects with which to investigate statistical 
turbulence theories which are not necessarily confined to 
reflection-symmetric configurations. All the usual issues such as scaling
exponent values, intermittency and anomaly could be studied for the
helical structure functions. This work provides a dimensional argument for 
what their scaling exponents could be. It is of interest to see how 
these behave relative to our predictions and to the anomalous scaling 
known for the usual structure functions. 

In conclusion, the present approach taken to derive the antisymmetric, or 
helical  K\'arm\'an-Howarth equation is not inconsistent with other recent 
work \cite{Chkh96,LPP97,GPP00}. However, our derivation
directly studies the dynamics of precisely those components of the
second-order correlation functions that were omitted in the
K\'arm\'an-Howarth equation because of assumed non-helicity of the
flow. The information about helicity of the flow is then obtained from
velocity correlations instead of velocity-vorticity correlations or
other correlations involving velocity derivatives. In a high-Reynolds
number experimental flow, measuring the second-order antisymmetric
velocity correlation function $R^A_{12}(r)$ in a coordinate system
chosen such that ${\bf r} = r {\bf {\hat k}}$, would give information
about the mean helicity in the flow, while measurement of
$S^A_{132}(r)$ would give information about the helicity flux. Such
objects are ideal candidates for detecting parity violation in flows
without having to resort to direct measurement of helicity.  We intend
to present the related  analysis of numerical and experimental
data in a future work.

\section{Acknowledgements}
I thank K.R. Sreenivasan for numerous discussions during my 
tenure as a graduate student at Yale University, which 
have lead to the present approach to the problem. 
D.D. Holm provided the impetus to consider the problem anew and valuable 
input during the work. I acknowledge useful discussions with Mark Taylor and 
Jamal Mohammed-Yusof. I am grateful to  G.L. Eyink and U. Frisch for their 
comments and criticisms, particularly concerning the issues of self-similarity
and scaling laws. 
The Erwin Schr\"odinger Institute of Mathematical Physics (Program 
on Developed Turbulence 2002), Vienna, provided partial support 
during this work.

\appendix
\section{Appendix}
The antisymmetric tensors $R^A_{\alpha \beta}$ and 
$S^A_{\alpha \gamma, \beta}$ are directly related to helicity 
dissipation and fluxes as we will now demonstrate.
Let us consider a particular geometry in which the separation vector $r$
is along the $z(3)$-axis. Then the only non-zero components are $R^A_{12} = 
-R^A_{21} = \langle u_1({\bf x})u_2({\bf x}+{\bf r}) - 
u_2({\bf x})u_1({\bf x}+{\bf r}) \rangle$. This particular object is 
the correlation of the two components of the velocity orthogonal to the 
vector ${\bf r}$. In the usual convention, it is the correlation of the
$v$ and $w$ components relative to the separation vector. The only 
non-zero contribution to the isotropic, antisymmetric velocity correlation
tensor comes from the velocity components orthogonal to the separation vector.

\begin{figure}
\begin{center}
\includegraphics[angle=0, width = 5cm, height = 4cm]{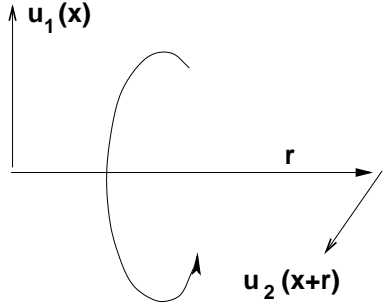}
\caption{Caricature of the type of correlation functions $R^A_{\alpha\beta}$ 
which are non-zero in flows that are not reflection-symmetric. The curved 
arrow indicates the ``handedness'' of the function.}
\label{cartoon}
\end{center}

\end{figure}

We contract the tensor $R^A_{\alpha \beta}$ with the antisymmetric 
tensor $\epsilon_{\alpha \beta \gamma}$ \cite{Holm_comm}.
\begin{eqnarray}
\epsilon_{\alpha \beta \gamma}R^A_{\alpha \beta}({\bf r}) &=& \epsilon_{\alpha \beta \gamma}H(r)\epsilon_{\alpha\beta\nu}{r_\nu \over
r}\nonumber\\ {\epsilon_{\alpha \beta \gamma}\langle u_\alpha({\bf x
}) u_\beta({\bf x} + {\bf r})\rangle +
\epsilon_{\beta\alpha\gamma}\langle u_\beta({\bf x}) u_\alpha({\bf x}+
{\bf r}) \rangle \over2} &=& 2 H(r) \delta_{\gamma\nu} {r_\nu \over r}
\nonumber \\
\langle {\bf u}({\bf x})\times {\bf u}({\bf x} + {\bf r})\rangle_\gamma
&=& 2 H(r){r_\gamma\over r}.
\label{rij_contract}
\end{eqnarray}
The (pseudo)scalar function $H(r)$ is the mean cross product of the 
velocities at
two points separated by the vector length scale ${\bf r}$. To choose 
a particular coordinate system, if the separation vector ${\bf r}$ lies along
the $z$-axis, then $H(r)$ is the $z$-component of the the cross-product. It 
vanishes as $|r| \rightarrow 0$. The physical picture is depicted in the 
cartoon of Fig. \ref{cartoon}.
This result is to be compared with the corresponding result for the
symmetric contribution $R^S_{\alpha \beta}$ contracted with
$\delta_{\alpha \beta}$. In that case, what is obtained is $\sim
\langle u_\alpha(x) u_\alpha(x+r) \rangle =
\langle {\bf u} ({\bf x}) \cdot {\bf u}({\bf x}+{\bf r})\rangle$, the
 mean scalar (dot) product of the velocities at two points
separated by the scale ${\bf r}$; as $r \rightarrow 0$, we recover the
non-zero mean energy $\sim \langle u^2 \rangle$. 

The function $R^A_{12}(r_3)$ may be thought of as a measure of
momentum transfer between two orthogonal components of velocity along
the direction perpendicular to both of them. If we take the curl of
$R^A_{\alpha \beta}$, we have
\begin{eqnarray}
\epsilon_{\alpha\beta\nu}\partial_\nu R^A_{\alpha \beta}({\bf r}) &=& \epsilon_{\alpha \beta \nu}\partial_\nu 
\Big( H(r) \epsilon_{\alpha \beta \gamma} {r_\gamma\over r}\Big) \nonumber\\
\langle {\bf u}({\bf x})\cdot {\bm \omega}({\bf x}+{\bf r})\rangle &=& 
2 {\partial H(r)\over \partial r} + 4 {H(r)\over r}.
\label{curl_rij}
\end{eqnarray}
Taking the limit as $r\rightarrow 0$, the left hand side reduces to
$\langle {\bf u}\cdot {\bm \omega}\rangle = 2 {\mathcal H}$ where
${\mathcal H}$ is the mean helicity of the flow, and we can solve for
what must be the leading order behavior of $H(r)$.
\begin{eqnarray}
H(r) = {1\over 3}{\mathcal H}r + \dots
\label{H_lead}
\end{eqnarray}
The scalar coefficient $H(r)$ of the antisymmetric tensor 
$R^A_{\alpha \beta}({\bf r})$ is, in the leading order, a direct measure of 
the mean helicity in the flow.
If we consider the particular coordinate system of 
Fig. \ref{cartoon} we see that $R^A_{12}(r) = H(r)$ which is a leading order 
measure of 
the mean helicity of the flow. We note again the advantage of this formulation 
which allows one to measure mean helicity using only velocity 
correlations, without having to measure any local gradients. 

We perform a similar analysis for the third-order object with
the contraction
\begin{eqnarray}
\epsilon_{\alpha \beta \mu}S_{\alpha \gamma \beta}^A =
&\epsilon_{\alpha \beta \mu}&{S(r)\over 2}\Big(2\epsilon_{\alpha \beta \nu}
{r_\nu r_\gamma\over r^2} + \epsilon_{\gamma \beta \nu}
{r_\nu r_\alpha \over r^2} - \epsilon_{\gamma\alpha\nu}
{r_\nu r_\beta \over r^2}\Big)\nonumber\\
\Big\langle u_\gamma(x)\Big( {\bf u}({\bf x}) \times {\bf u}({\bf x}+{\bf r})\Big)_\mu \Big\rangle &=& {S(r)}\Big(3 {r_\gamma r_\mu\over r^2} -\delta_{\gamma\mu}\Big)
\label{almost_LPP}
\end{eqnarray}
If we now proceed to write, as in \cite{LPP97}, the velocity vector as
the sum of its longitudinal (along ${\bf r}$) and transverse
components such that ${\bf u}({\bf x}) = {\bf u}_L({\bf x}) + {\bf
u}_T({\bf x})$, we have
\begin{eqnarray}
\Big\langle (u_L + u_T)_\gamma \Big( ({\bf u}_L + {\bf u}_T) \times 
({\bf u}'_L + {\bf u}'_T)\Big)_\mu \Big\rangle &=& {S(r)}\Big(3
{r_\gamma r_\mu\over r^2} -\delta_{\gamma\mu}\Big)
\end{eqnarray}
where the un-primed velocities denote measurement at ${\bf x}$ and the 
primed velocities denote measurement at ${\bf x} + {\bf r}$. It is clear that
both the left and right side vanish for $\gamma = \mu$. But we would like to
examine the detailed balance in terms of the longitudinal and transverse 
components on the left hand side. 
\begin{eqnarray}
M_{\gamma\mu} = \Big\langle (u_L + u_T)_\gamma &\Big(& ({\bf u}_L \times
{\bf u}'_L)_\mu + ({\bf u}_L \times {\bf u}'_T)_\mu + ({\bf u}_T
\times {\bf u}'_L)_\mu + ({\bf u}_T \times {\bf u}'_T)_\mu
\Big)\Big\rangle \nonumber\\ 
 &&= S(r)\Big(3{r_\gamma r_\mu\over r^2} -\delta_{\gamma\mu}\Big).
\label{LPP_wrong} 
\end{eqnarray}
Since this must be true for any choice ${\bf r}$, we can, 
without loss of generality, choose ${\bf r} = r {\bf \hat i}$ so that it lies
along the $x$-axis. The matrix of Eq. (\ref{LPP_wrong}) is then diagonal 
and traceless, and we see, using Eq. \ref{rsq}, that
\begin{eqnarray}
M_{11} &=& {\bf u}_L\cdot ({\bf u}_T \times {\bf u}'_T) = 2S(r) = {1\over
15}h r^2 \label{LPP_correct}\\
M_{22} &=& M_{33}  = {\bf u}_T \cdot (({\bf u}_L + {\bf u}_T) 
\times ({\bf u}'_L+{\bf u}'_T)) = -{S(r)} = -{1\over 30}h r^2
\nonumber
\end{eqnarray}
Eq. (\ref{LPP_correct}) is a form of the so-called ``2/15ths law'' derived
by L'vov et al \cite{LPP97}. The exact result of that work is
obtained by computing $({\bf u}_L - {\bf u}'_L)\cdot ({\bf u}_T \times
{\bf u}'_T))$ which is equal, by homogeneity to $2{\bf u}_L\cdot ({\bf
u}_T \times {\bf u}'_T) = 4 S(r) = {2 \over 15}h r^2$.
	
\end{spacing}

\end{document}